\documentclass[12pt,a4paper]{iopart}
\usepackage{iopams}  

\usepackage{graphicx}
\usepackage{float} 
\usepackage{subfig} 

\usepackage{ifpdf}
\ifpdf
\usepackage[pdftex]{hyperref}
\else
\usepackage[hypertex]{hyperref}
\fi

\hypersetup{
  pdftitle={},%
  pdfauthor={},%
  pdfsubject={},%
  pdfkeywords={},%
  pdfstartview={},%
  bookmarksopen=true, breaklinks=true, debug=true, %
  colorlinks=true, linkcolor=red, citecolor=blue, urlcolor=blue
}

\newcommand{\beq}{\begin{eqnarray}}
\newcommand{\eeq}{\end{eqnarray}}
\newcommand{\be}{\begin{eqnarray*}}
\newcommand{\ee}{\end{eqnarray*}}

\newcommand{\ie}{{\it i.e.}}

\newcommand{\cf}[1]{{Fig.~\ref{#1}}}
\newcommand{\ct}[1]{{Table.~\ref{#1}}}

\def\lsim{\raise0.3ex\hbox{$<$\kern-0.75em\raise-1.1ex\hbox{$\sim$}}}
\def\gsim{\raise0.3ex\hbox{$>$\kern-0.75em\raise-1.1ex\hbox{$\sim$}}}

\def\CuCu {CuCu}

\def\dAu  {$d$Au}
\def\dAum  {d\mathrm{Au}}
\def\AuAu {AuAu}
\def\pp   {$pp$}
\def\pA   {$pA$}
\def\dA {$dA$}
\def\AA   {$AA$}
\def\AB   {$AB$}

\def\sqrtsNN {\mbox{$\sqrt{s_{NN}}$}}

\def\Ncoll   {\mbox{$N_{\rm coll}$}}

\def\RdAu    {\mbox{$R_{d\rm Au}$}}

\def\jpsi   {\mbox{$J/\psi$}}
\def\ccbar {\mbox{$c\bar{c}$}}
\def\pT      {\mbox{$P_{T}$}}
\def\kT     {\mbox{$k_{T}$}}
\def\sigabs {\mbox{$\sigma_{\mathrm{abs}}$}}
\def\beq     {\begin{equation}}
\def\eeq     {\end{equation}}

\long\def\symbolfootnote[#1]#2{\begingroup%
  \def\thefootnote{\fnsymbol{footnote}}\footnote[#1]{#2}\endgroup}

\graphicspath{{plots/}}

\begin{document}

%
\title[Uncertainties in the extraction of $\sigma_{\rm abs}^{J/\psi}$ in cold nuclear matter]{On the theoretical and experimental uncertainties in the extraction of the \jpsi\ absorption cross section in cold nuclear matter}

\author[ A.~Rakotozafindrabe, E.~G.~Ferreiro, F.~Fleuret and J.~P.~Lansberg]{A.~Rakotozafindrabe$^{\dag}$, E.~G.~Ferreiro$^{\ddag}$, F.~Fleuret$^{\S}$ and J.~P.~Lansberg$^{\P}$}

\address{$^\dag$ IRFU/SPhN, CEA Saclay, 91191 Gif-sur-Yvette Cedex, France}
\address{$^\ddag$ Departamento de F{\'\i}sica de Part{\'\i}culas, 
Universidad de Santiago de Compostela, 15782 Santiago de Compostela, Spain}
\address{$^\S$ Laboratoire Leprince Ringuet, \'Ecole polytechnique, CNRS-IN2P3,  91128 Palaiseau, France}
\address{$^\P$ Centre de Physique Th\'eorique, \'Ecole Polytechnique,
  CNRS, 91128 Palaiseau, France}
\ead{andry.rakotozafindrabe@cea.fr}

\author{}

\address{}

\begin{abstract}
We investigate the cold nuclear matter effects on \jpsi\ production,
whose understanding is fundamental to study the quark-gluon plasma. Two of these effects are of particular
relevance: the shadowing of the parton distributions and the nuclear absorption of the
\ccbar~pair. If \jpsi's are not produced {\it via} a $2 \to 1$ process as suggested by recent theoretical works, one has to modify accordingly the way to compute the nuclear
shadowing. This naturally induces differences in the absorption cross-section fit to the data. A careful analysis of these differences however requires taking into account the experimental uncertainties and their correlations, as done in this work for \dAu\ collisions at $\sqrtsNN=200\mathrm{~GeV}$, using several shadowing parametrisations.
\end{abstract}

\vspace{2pc}


\section{Introduction}
\label{sec:intro}

Relativistic nucleus-nucleus (\AB) collisions are expected to produce a deconfined state of QCD matter -- the Quark Gluon Plasma (QGP) -- at high enough densities or temperatures. It has long been suggested~\cite{Matsui86} that the \jpsi\ meson would be sensitive to Hot and Dense Matter (HDM) effects, through mechanisms like the dissociation of the \ccbar~pair due to the colour Debye screening. A significant suppression of the \jpsi\ yield was observed by the PHENIX experiment in \CuCu~\cite{Adare:2008sh} and \AuAu~\cite{Adare:2006ns} collisions at $\sqrtsNN=200\mathrm{~GeV}$. However, before giving any interpretation, Cold Nuclear Matter~(CNM) effects have to be properly disentangled and subtracted. They are known to impact the \jpsi\ production in proton-nucleus (\pA) or deuteron-nucleus (\dA) collisions, where the deconfinement conditions can not be reached. Such non-trivial effects are demonstrated by the PHENIX \dAu\ data~\cite{Adare:2007gn} obtained at the same energy. Two CNM effects are of particular importance~\cite{Vogt:2004dh}: (i)  the shadowing of the initial parton distributions (PDFs)
due to the nuclear environment, and (ii) the breakup of \ccbar~pairs
consecutive to multiple scatterings with the remnants of the
projectile and target nuclei, referred to as the nuclear
absorption. Recent theoretical
works incorporating QCD corrections or $s$-channel cut contributions have emphasized~\cite{Haberzettl:2007kj,QCD:recentWorks} that the Colour-Singlet (CS) mediated contributions are sufficient to describe the experimental data for hadroproduction of both charmonium and bottomonium systems without the need of Colour-Octet (CO) contributions. Furthermore, recent works~\cite{ee} focusing on production at $e^+ e^-$ colliders have posed stringent constraints on the size of CO contributions, which are the precise ones supporting a $2 \to 1$ hadroproduction mechanism~\cite{Lansberg:2006dh}. As a consequence, \jpsi\ production at low and mid \pT\ likely proceeds via a $2 \to 2$ process, such as $g+g \to \jpsi + g$, instead of a $2 \to 1$ process. As we have shown
in previous
studies~\cite{Ferreiro:2008wc,Ferreiro:2009qr,Ferreiro:2009ur}, this
modifies both the way to compute the nuclear shadowing and its
expected impact on the \jpsi\ production. In this work, we shall focus
on the changes induced on the rapidity dependence of the \jpsi\
nuclear modification factor in \dAu, while using several
parametrisations of the nuclear PDF. As in~\cite{Adare:2007gn} --
where CNM effects were computed based on a $2 \to 1$ kinematics -- we
shall use the same \dAu\ data to derive the absorption
cross-section \sigabs\ required on top of the shadowing, but assuming here a $2
\to 2$~underlying partonic process. We shall compare the results found
in both schemes, with a special emphasis on the limitations from the
experimental uncertainties.

The article is organized as follows. In section~\ref{sec:approach},
we will describe our model and the method chosen to carry a
data-driven evaluation of the nuclear absorption cross-section. And in
section~\ref{sec:results}, we will present and discuss our
results before concluding.

\section{Our approach}
\label{sec:approach}

To describe the \jpsi\ production in nucleus collisions, our
Monte~Carlo framework~\cite{Ferreiro:2008wc,OurIntrinsicPaper} is based on the probabilistic Glauber model, the
nuclear density profiles being defined with the Woods-Saxon
parameterisation for any nucleus ${A>2}$ and the Hulthen wavefunction
for the deuteron~\cite{Hodgson:1971}. The nucleon-nucleon inelastic cross section at
$\sqrtsNN=200\mathrm{~GeV}$ is taken to $\sigma_{NN}=42\mathrm{~mb}$ and the
maximum nucleon density to $\rho_0=0.17\mathrm{~nucleons/fm}^3$.

In order to study the \jpsi\ production, we need to implement
in our Monte Carlo the following ingredients: the partonic process for the
\ccbar\ production and the CNM effects.

\subsection{Partonic process for the \ccbar\ production}
\label{subsec:partonic process}

Most studies on the \jpsi\ production in hadronic collisions rely on the assumption that
the \ccbar~pair is produced by the fusion of two gluons carrying
some intrinsic transverse momentum~\kT. The partonic process being a
\mbox{$2\to 1$} scattering, the sum of the gluon intrinsic transverse momentum is transferred to the \ccbar~pair, thus to
the \jpsi\ since the soft hadronisation process does not alter
significantly the
kinematics. This is supported by the picture of the Colour Evaporation
Model (CEM) at LO (see~\cite{Lansberg:2006dh} and references 
therein) or of the CO mechanism at
$\alpha_s^2$~\cite{Cho:1995ce}. In such approaches, the transverse momentum \pT\ of the
\jpsi\ {\it entirely} comes from the intrinsic transverse momentum of the initial gluons.

However, the average value of~\kT\ is not expected to go much beyond
$\sim 1\mathrm{~GeV}$. So this process is not sufficient to describe the \pT\ spectrum of quarkonia produced in
hadron collisions~\cite{Lansberg:2006dh}. For $\pT \, \gsim
\,2-3\mathrm{~GeV}$, most of the transverse momentum should have an extrinsic
origin, \ie\ the \jpsi's \pT\ would be balanced by the emission of a recoiling particle in the final
state. 
The \jpsi\ would then be produced by gluon fusion in a \mbox{$2\to 2$} process with emission of a hard final-state gluon.
This emission, which is anyhow mandatory to conserve $C$-parity, 
has a definite influence on the kinematics of the
\jpsi\ production. Indeed, for a given \jpsi\ momentum (thus for
fixed rapidity~$y$ and \pT), the processes discussed above, \ie\ $g+g \to \ccbar \to J/\psi \,(+X)$
and $g+g \to J/\psi +g$,  will proceed on the average from initial gluons with different Bjorken-$x$. Therefore,
 they will be affected by different shadowing corrections. From now on, we will refer to the former
scenario as the {\it intrinsic} scheme, and to the latter as the {\it extrinsic} scheme.

In the intrinsic scheme, we use the fits to the $y$ and \pT\ spectra measured
by PHENIX~\cite{Adare:2006kf} in \pp\ collisions at $\sqrt{s_{NN}}=200\mathrm{~GeV}$
as inputs of the Monte-Carlo.
Indeed, the measurement of the \jpsi\ momentum completely fixes the longitudinal
 momentum fraction carried by the initial partons:
\begin{equation}
x_{1,2} = \frac{m_T}{\sqrt{s_{NN}}} \exp{(\pm y)} \equiv x_{1,2}^0(y,P_T),
\label{eq:intr-x1-x2-expr}
\end{equation}
with the transverse mass $m_T=\sqrt{M^2+P_T^2}$, $M$ being the \jpsi\ mass.

On the other hand, in the extrinsic scheme, information from the data alone
-- the $y$ and \pT\ spectra -- is not sufficient to determine $x_1$ and $x_2$.
Actually, the presence of a final-state gluon introduces further degrees
of freedom in the kinematics, allowing several $(x_1, x_2)$ for a given set $(y, P_T)$.
 The four-momentum conservation explicitely results in a more complex expression of $x_2$ as a function of~$(x_1,y,P_T)$:
\begin{equation}
x_2 = \frac{ x_1 m_T \sqrt{s_{NN}} e^{-y}-M^2 }
{ \sqrt{s_{NN}} ( \sqrt{s_{NN}}x_1 - m_T e^{y})} \ .
\label{eq:x2-extrinsic}
\end{equation}
Equivalently, a similar expression can be written for $x_1$ as a function of~$(x_2,y,P_T)$.
Even if the kinematics determines
the physical phase space, models are anyhow {\it mandatory} to compute the proper
weighting of each kinematically allowed $(x_1, x_2)$. This weight is simply
the differential cross section at the partonic level times the gluon PDFs,
\ie\ $g(x_1,\mu_F) g(x_2, \mu_F) \, d\sigma_{gg\to J/\psi + g} /dy \,
dP_T\, dx_1 dx_2 $.
In the present implementation of our code, we are able to use the partonic differential
cross section computed from {\it any} theoretical approach. For now, we use the one
from~\cite{Haberzettl:2007kj} which takes into account the $s$-channel cut
contributions~\cite{Lansberg:2005pc} to the basic CS model and
satisfactorily describes the PHENIX pp data~\cite{Adare:2006kf} down to very low~\pT~\cite{Ferreiro:2008wc}. 

\subsection{Shadowing and nuclear absorption}
\label{subsec:shadowing-absorption}

To obtain the \jpsi\ yield in \pA\ and \AA\ collisions, a shadowing-correction
factor has to be applied to the \jpsi\ yield obtained from the simple
superposition of the equivalent number of \pp\ collisions.
This shadowing factor can be expressed in terms of the ratios $R_i^A$ of the
nuclear Parton Distribution Functions (nPDF) in a nucleon belonging to a nucleus~$A$ to the
PDF in the free nucleon:
 
\begin{equation}
\label{eq:shadow-corr-factor}
R^A_i (x,Q^2) = \frac{f^A_i (x,Q^2)}{ A f^{nucleon}_i (x,Q^2)}\ , \ \
i = q, \bar{q}, g \ .
\end{equation}
The numerical parameterisation of $R_i^A(x,Q^2)$
is given for all parton flavours. Here, we restrict our study to gluons since, at
high energy, the \jpsi\ is essentially produced through gluon fusion
\cite{Lansberg:2006dh}. In order to see how the CNM effects can vary
depending on the different shadowing parametrisations used as an input, we will consider three
of them: nDSg~\cite{deFlorian:2003qf} at LO, EKS98~\cite{Eskola:1998df}  and
EPS08~\cite{Eskola:2008ca}. They span the current evaluation of the
uncertainty~\cite{Eskola:2009uj} on the gluon nPDF, from a small to a very large antishadowing.

The second CNM effect that we are going to take into account concerns
the nuclear absorption.  In the framework of the probabilistic Glauber
model, this effect is usually parametrised
by introducing an effective absorption cross
section~\sigabs\ of the pre-resonant \ccbar~pair
when propagating in the nuclear medium.
In the following, we shall compare the data-constrained
values of \sigabs\ given two different partonic \ccbar\ production mechanisms --~intrinsic and
extrinsic -- and three shadowing parametrisations as cited above.

\subsection{A data-driven evaluation of the nuclear absorption cross-section}
\label{subsec:fit-procedure}

We present here the derivation of the \sigabs\ values consistent with the PHENIX \dAu\ data, taking into account the various experimental uncertainties and their correlations.
To do so, we have chosen to follow the procedure from~\cite{Adare:2007gn,Adare:2008cg}. For a given choice of the \ccbar\ production mechanism and nPDF parametrisation, the best possible agreement of the theory to the data is obtained for the value of \sigabs\ that minimizes the quantity:

\begin{equation}
\label{eq:chi-2}
\chi^2 (\vec{p}, \epsilon_b, \epsilon_c)= \sum_{i=1}^n \frac{\left( d_i + \epsilon_b \sigma_{b_i} + \epsilon_c d_i \sigma_c - \mu_i(\vec{p}\,) \right)^2}{ \tilde{\sigma}_i^2 } + \epsilon_b^2 + \epsilon_c^2 \, 
\end{equation}
with $\tilde{\sigma}_i = \sigma_i \left(  d_i + \epsilon_b \sigma_{b_i} + \epsilon_c d_i \sigma_c \right)$, $d_i$ being the set of experimental values from PHENIX, $\mu_i$~the respective values predicted by the theory for a given set of parameters $\vec{p}$ (\ie\ the nPDF parametrisation and \sigabs), $\sigma_i$ are the point-to-point uncorrelated errors (statistical and systematic), $\sigma_{b_i}$ the point-to-point correlated systematic errors, $\sigma_c$ the global systematic error on the normalisation of the data and $\epsilon_{(b,c)}$ the fractions of the systematic uncertainties $\sigma_{(b_i,c)}$ used to shift the data points~$d_i$.

\section{Results and discussion}
\label{sec:results}

In the following, we present our results for the \jpsi\ nuclear modification factor in \dAu\ collisions: $R_{\dAum} = dN_{\dAum}^{J/\psi}/\langle\Ncoll\rangle dN_{pp}^{J/\psi}$,
where $dN_{\dAum}^{J/\psi} (dN_{pp}^{J/\psi})$ is the observed \jpsi\ yield in \dAu\ (\pp) collisions
and $\langle\Ncoll\rangle$ is the average number of nucleon-nucleon collisions occurring
in one \dAu\ collision. Without nuclear effects, $R_{\dAum}$ should equal unity.

\begin{table}[hbt!]
\begin{center}
\begin{tabular}{c|cc||c|cc}
\hline
& $\sigma_{\mathrm{abs}}$ (mb) & $\chi^2_{min}$ & & $\sigma_{\mathrm{abs}}$ (mb) & $\chi^2_{min}$ \\ 
\hline\hline 
nDSg  Int. & $2.2^{+2.6}_{-2.2}$ & 1.6 & nDSg  Ext. & $3.0^{+2.5}_{-2.4}$ & 1.4\\
EKS98 Int. & $3.2 \pm 2.4$ & 0.9 & EKS98 Ext. & $3.9^{+2.7}_{-2.3}$ & 1.1 \\
EPS08 Int. & $2.1^{+2.6}_{-2.2}$ & 1.1 & EPS08 Ext. & $3.6^{+2.4}_{-2.5}$ & 0.5 \\
\hline 
\end{tabular}
\end{center}
\caption{\sigabs\ extracted from fits of \RdAu\ vs y for the intrinsic (left) and extrinsic (right) schemes, when considering all the different
  types of errors on the data, together with the corresponding $\chi^2$ obtained for the best fit.}
\label{tab:sigabs-values}
\end{table}

In \ct{tab:sigabs-values}, we recall the results from~\cite{Ferreiro:2009ur}, with the value of \sigabs\ corresponding to the best fit to PHENIX data \RdAu\ vs~$y$, and the one standard deviation uncertainties. As in~\cite{Adare:2007gn}, this extraction relies on the assumption that \sigabs\ is independent of~$y$. The best agreement to the data is obtained in the extrinsic scheme with EPS08. The larger is the antishadowing, the larger are the differences between both schemes. This is visible in the $\chi^2$ for the best fit or on the different shapes of the curves in \cf{fig:RdAu-vs-y-int-ext} which shows the obtained CNM effects together with PHENIX data for \RdAu\ vs~$y$. 
In this plot, the anti-shadowing peak in \RdAu\  is systematically shifted towards larger~$y$ for the extrinsic scheme with respect to the one in the intrinsic case. This reflects the larger value of the gluon momentum fraction $x_2$ in the Au nucleus needed to produce a \jpsi\ when the momentum of the final state gluon is indeed accounted for. We shall now focus on the bands picturing the CNM effects in both schemes which account for the experimental uncertainties. For the time being, the current size of the errors unfortunately does not allow to distinguish between the two approaches if assuming a constant \sigabs\ with~$y$.

Using the same procedure, we have qualitatively studied three scenarios with thirty times more statistics and a) no improvement, b) a reduction of $35 \%$  and c) a reduction of $50\%$ of the systematics. We have found that, with the same assumption on \sigabs\ (namely constant vs~$y$), the latest \dAu\ data (2008) will not be sufficient to distinguish between a $2\to 1$ and a $2\to 2$ production mechanism by {\it only} looking at the $y$-dependence of \RdAu, unless the (anti)shadowing is as strong as encoded in the EPS08 parametrisation.

\begin{figure*}[htb!]
\begin{center}
\hspace{-0.84cm}
\subfloat[][nDSg at LO.]{%
\label{fig:RdAu-vs-y-int-ext-a}
\includegraphics[width=.34\linewidth]{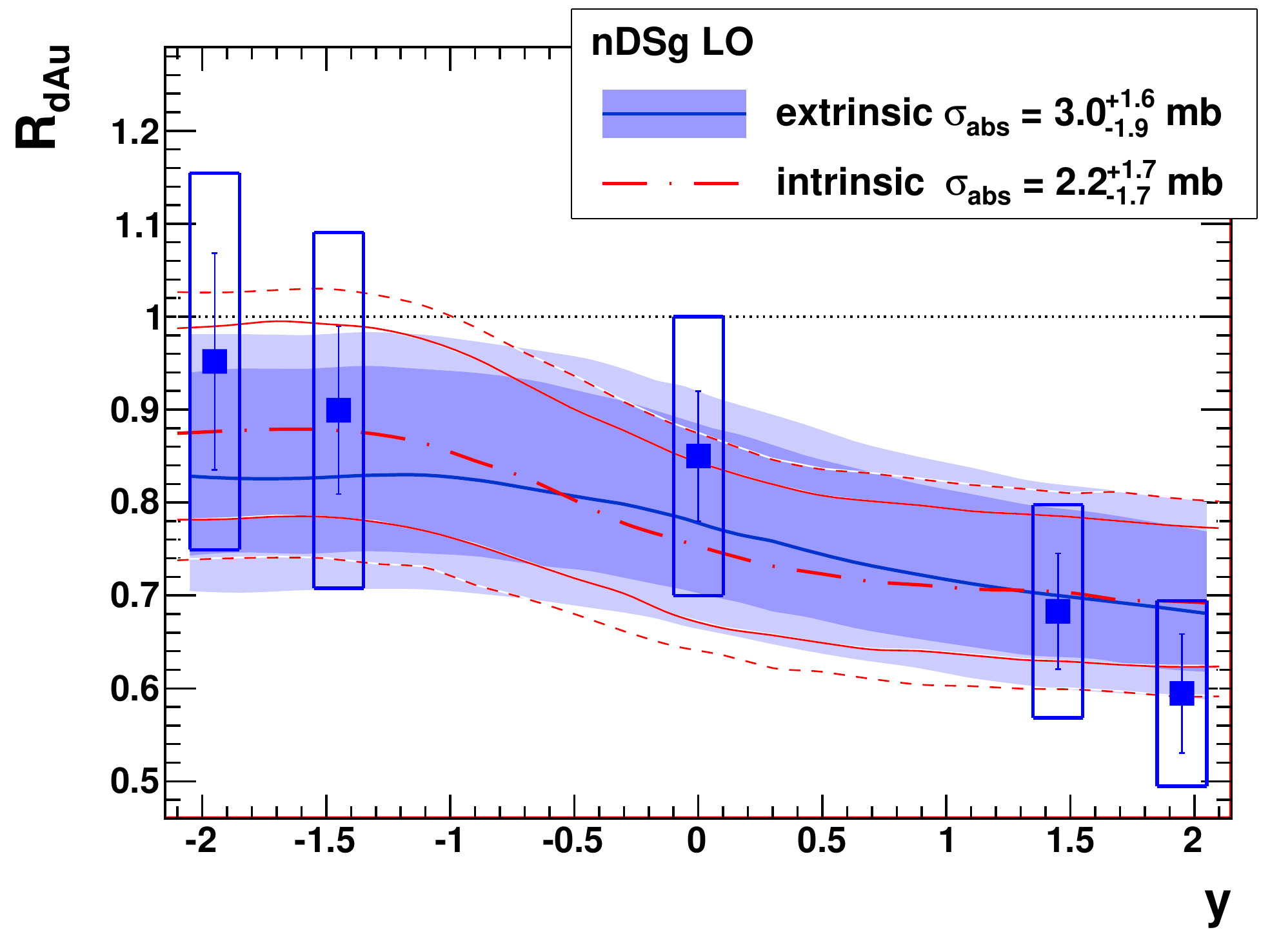}}
\hspace{-0.33cm}
\subfloat[][EKS98.]{%
\label{fig:RdAu-vs-y-int-ext-b}
\includegraphics[width=.34\linewidth]{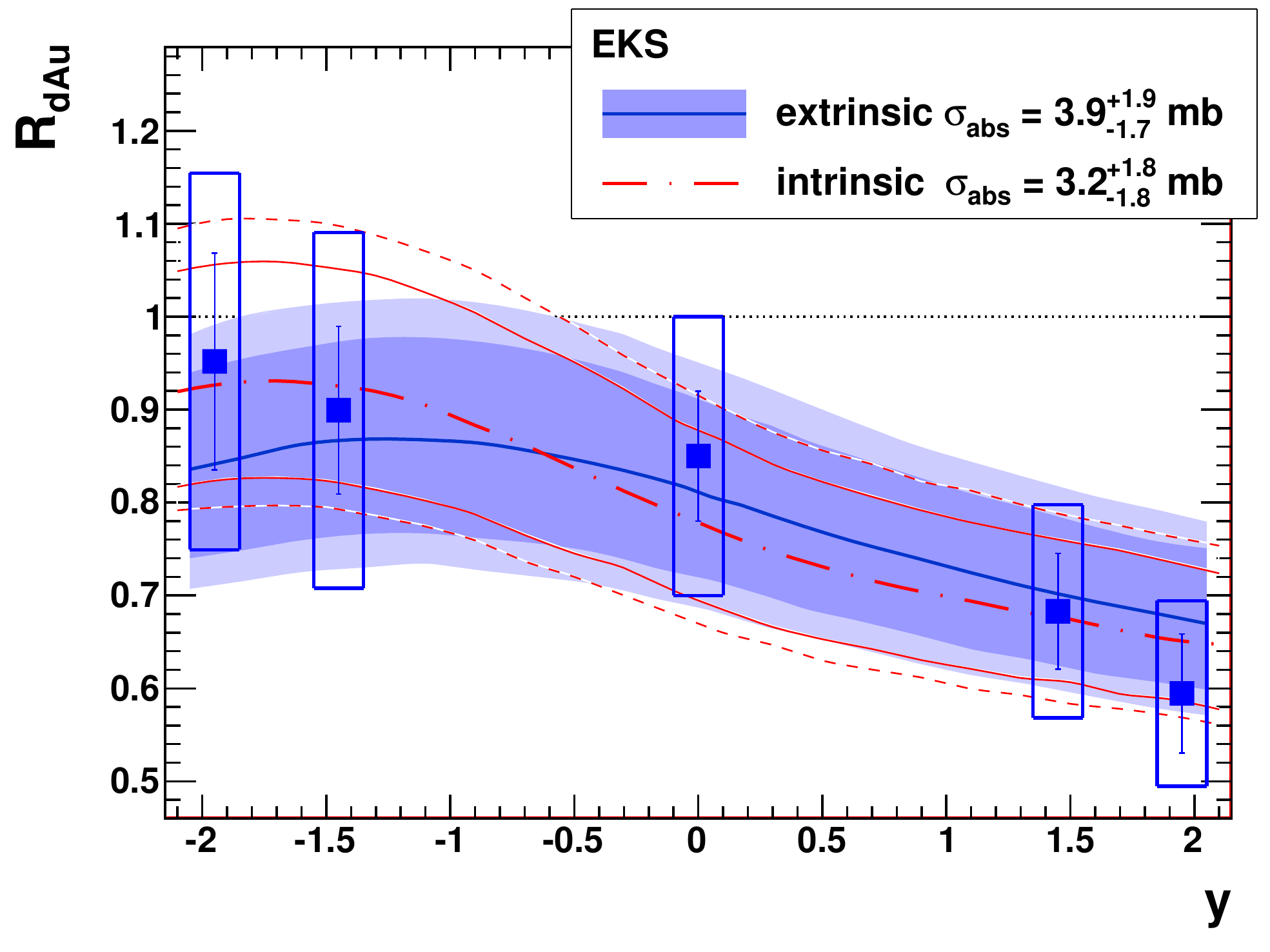}}
\hspace{-0.33cm}
\subfloat[][EPS08.]{%
\label{fig:RdAu-vs-y-int-ext-c}
\includegraphics[width=.34\linewidth]{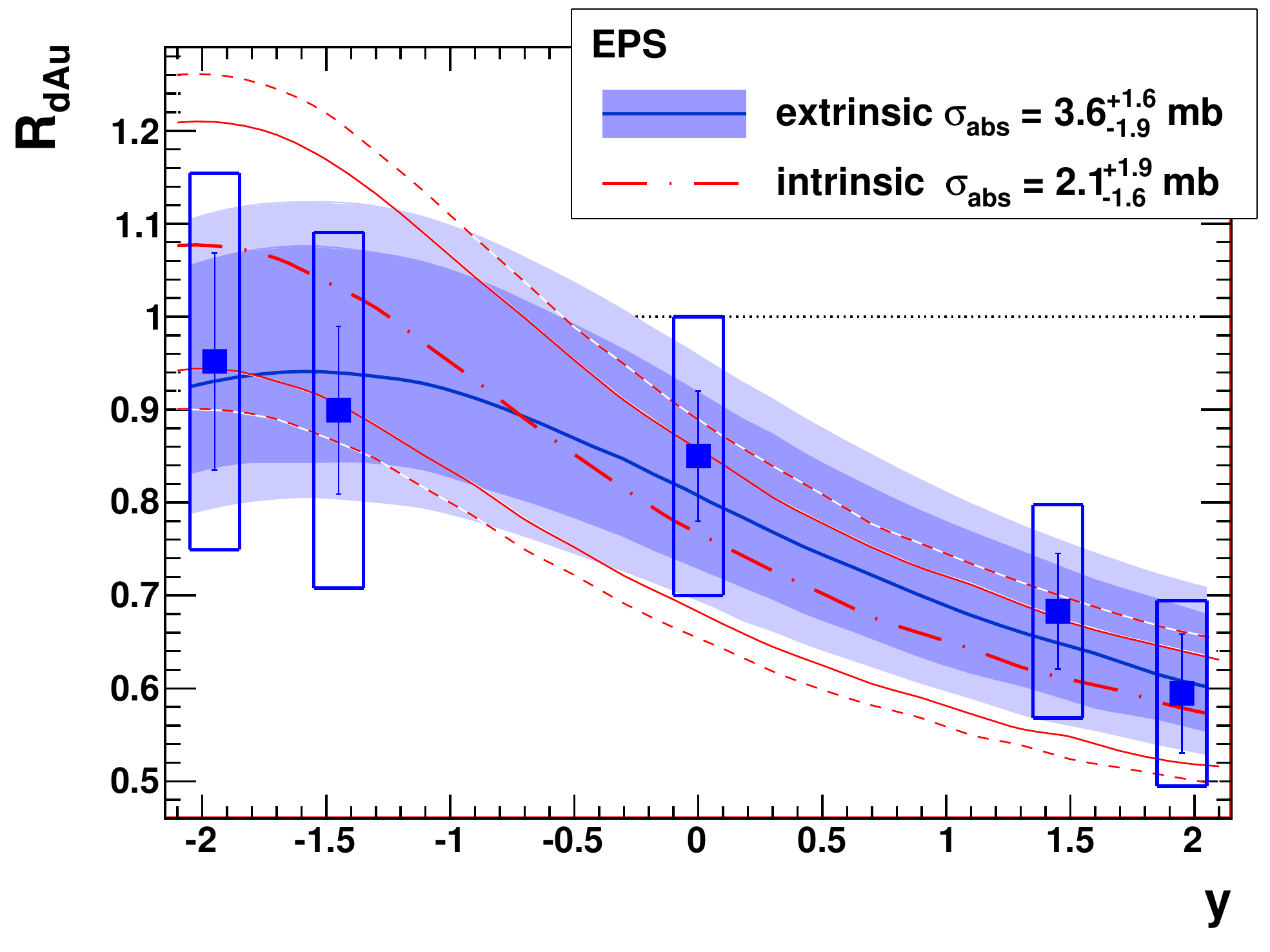}}
\end{center}
\caption{\RdAu\ vs $y$ for the intrinsic
  and extrinsic schemes and for the three nPDF parametrisations, compared to PHENIX data~\cite{Adare:2007gn}. For each scheme, the central band represents
  the range in \sigabs\ consistent with the
  data within one standard deviation, when taking into account the
  point-to-point uncorrelated (bar) and correlated (box) errors. The corresponding \sigabs\
  values are reported in the legend. The outer band is the
  obtained range in \sigabs\ when
  the global error on the data normalisation is also considered.}
\label{fig:RdAu-vs-y-int-ext}
\end{figure*}


\section*{Acknowledgments}

We are indebted to M.\ J.\ Leitch for helping us to implement the procedure for the $\chi^2$ minimisation.

\end{document}